\documentclass[preprint,12pt]{elsarticle}
\usepackage{amssymb}
\biboptions{numbers}

\journal{Optics Communications}

\begin{document}

\begin{frontmatter}

\title{Nanofiber-Based Double-Helix Dipole Trap for Cold Neutral Atoms}

\author{Daniel Reitz}
\ead{dreitz@ati.ac.at}
\author{Arno Rauschenbeutel\corref{corrAuthor}}
\ead{arno.rauschenbeutel@ati.ac.at}
\address{Vienna Center for Quantum Science and Technology, Atominstitut, TU Wien, Stadionallee 2, 1020 Wien, Austria}
\cortext[corrAuthor]{Corresponding author}

\begin{abstract} A double-helix optical trapping potential for cold atoms can be straightforwardly created inside the evanescent field of an optical nanofiber. It suffices to send three circularly polarized light fields through the nanofiber; two counterpropagating and far red-detuned with respect to the atomic transition and the third far blue-detuned. Assuming realistic experimental parameters, the transverse confinement of the resulting potential allows one to reach the one-dimensional regime with cesium atoms for temperatures of several $\mu$K. Moreover, by locally varying the nanofiber diameter, the radius and pitch of the double-helix can be modulated, thereby opening a realm of applications in cold-atom physics.\end{abstract}


\end{frontmatter}

In recent experiments, we have demonstrated that laser-cooled cesium atoms can be trapped using a two-color evanescent field surrounding the nanofiber waist of a tapered optical fiber (TOF) \cite{Vetsch2010}. There, the atoms were trapped in a one-dimensional optical lattice, 200~nm above the nanofiber surface. This trapping configuration was achieved by appropriately choosing wavelength, polarization, and optical power of the trapping field, and by interfering counterpropagating modes \cite{LeKien2004b}. As an important asset, the method allows one to efficiently interface the atoms with near-resonant light sent through the TOF \cite{Vetsch2010,Dawkins2011}. 

Furthermore, in former theoretical work, we have shown that three-dimen\-sional trapping of cold atoms in the evanescent field surrounding optical nanofibers is also possible with two co-propagating nanofiber-guided modes of the same wavelength when resorting to the interference of higher order transverse modes \cite{Sague2008}. This already pointed towards the possibility of realizing extraordinary trapping configurations by means of light fields guided in optical nanofibers. 

Here, we show that a double-helix optical trapping potential for cold atoms can be straightforwardly created using a similar optical set-up as in \cite{Vetsch2010}. Our approach is an advantageous alternative to the generation of helical potentials with free space optics using, e.g., the interference of co-propagating Laguerre-Gaussian and Gaussian beams \cite{MacDonald2002}. In particular, it is not limited by the Rayleigh length, neither concerning the maximum length over which the helix potential can be made uniform nor concerning the minimum length over which the helix parameters can be modulated.

Helical trapping potentials, in which particles are confined in two spatial dimensions and are freely moving along the helix, can lead to interesting physical effects \cite{Exner052007,Bhattacharya072007,Law082008,Schmelcher092011,Okulov112011}. In particular, if a long ranged interaction is present, like for polar molecules \cite{Law082008} or for charged particles \cite{Schmelcher092011,Kibis051992}, bound states with a fixed interparticle distance can occur. In this situation, it should, e.g., become possible to study zero-temperature second-order liquid-gas transitions \cite{Law082008}. Moreover, if the radius and pitch of the helical potential exhibit local variations, geometrically induced bound states will appear. Their origin lies in the fact that the curvature of an otherwise constant waveguide for quantum particles leads to the presence of an attractive potential \cite{Exner052007,Bhattacharya072007,Exner111989}. Such effects are, e.g., relevant for electron transport through quantum wires in quantum heterostructures \cite{Londergan1999}. Studying the underlying physics in a clean and configurable model system using cold atoms might therefore lead to valuable insights.

\begin{figure}
\centerline{\includegraphics[width=8cm]{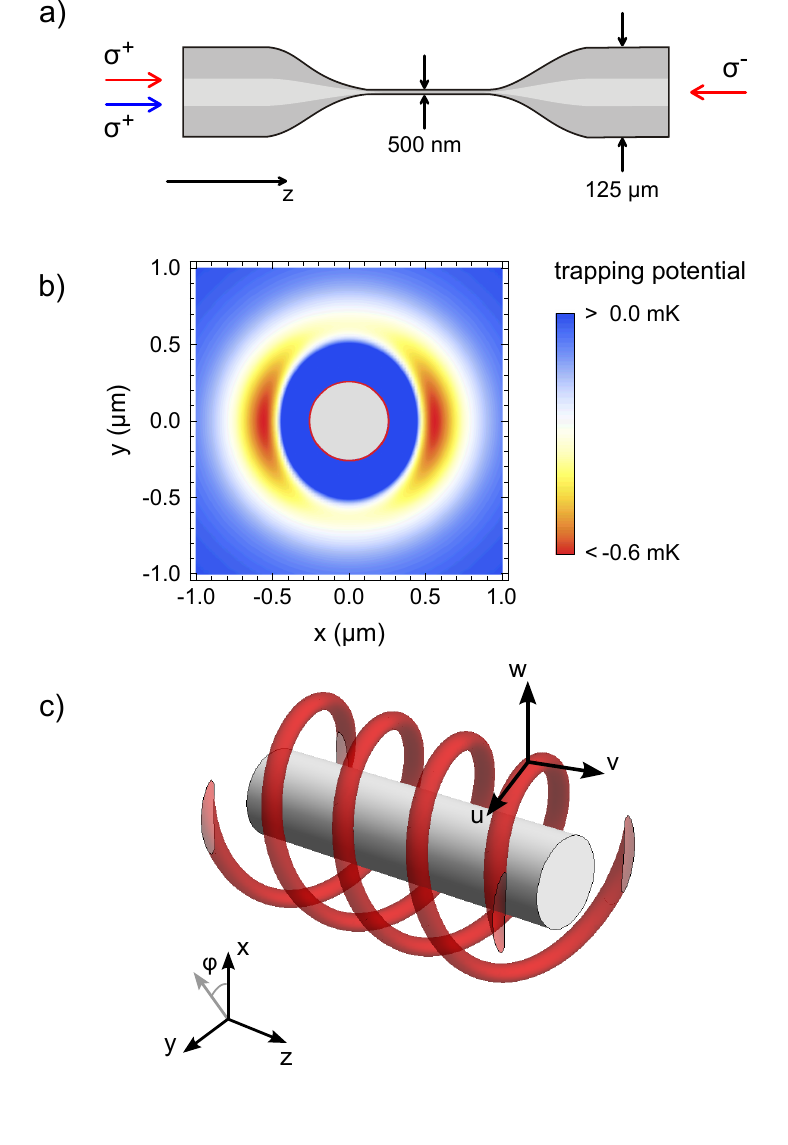}}
\caption{(a) Schematic of the optical set-up of the nanofiber-based double-helix trap, see text for details. (b) Plot of the resulting double-helix trapping potential for cesium atoms around a nanofiber with a diameter of 500~nm. The plane is chosen perpendicular to the nanofiber axis at $z=0$. See text for the wavelengths and powers of the trapping fields. (c) Three-dimensional plot of the same double-helix trapping potential. The helical tubes correspond to equipotential surfaces $k_{\rm B}\cdot 50~\mu$K above the trapping minimum, where $k_{\rm B}$ is Boltzmann's constant. Each helix has a radius of $R\approx 560$~nm and a pitch of $h\approx 1~\mu$m.
}\label{Fig1}
\end{figure}
Figure~\ref{Fig1}(a) schematically shows the TOF with the nanofiber waist and the trapping fields. The red-detuned fields are launched from both sides into the TOF with $\sigma^+$ ($\sigma^-$) polarization for the field propagating in the positive (negative) $z$-direction. The blue-detuned field is $\sigma^+$ polarized and propagates in the positive $z$-direction only. In the taper transitions of the TOF, the circularly polarized LP$_{01}$ modes of the unprocessed fiber ends of the TOF \cite{Yariv1991} are adiabatically transformed into superpositions of the quasi-linearly polarized fundamental HE$_{11}$ modes of the nanofiber waist \cite{LeKien2004} which are described by the normalized mode functions
\begin{equation}\label{eq:linpolHE11}
\begin{array}{l}
{\bf E}^{\pm}_x (r, \varphi, z) = [{\bf e}_x (r, \varphi) \pm i {\bf e}_z (r) \cos \varphi]e^{\mp i\beta z}\\
{\bf E}^{\pm}_y (r, \varphi, z) = [{\bf e}_y (r, \varphi) \pm i {\bf e}_z (r) \sin \varphi]e^{\mp i\beta z}.
\end{array}
\end{equation}
Here, the quasi-linear polarization is along $x$ and $y$, respectively, the index $+$ ($-$) stands for propagation in the positive (negative) $z$-direction, ${\bf e}_x (r, \varphi)$ and ${\bf e}_y (r, \varphi)$ are the real-valued polarization vector fields corresponding to the transverse components of the modes, ${\bf e}_z (r)\cos \varphi$ and ${\bf e}_z (r)\sin \varphi$ are the real-valued polarization vector fields corresponding to the longitudinal component of the modes, and $\beta$ is the wavelength-dependent propagation constant.

The counterpropagating quasi-circularly polarized nanofiber modes are thus given by
\begin{equation}\label{eq:circpolHE11}
\begin{array}{l}
{\bf E}^{\rm +}_{\sigma^+} (r, \varphi, z)=\frac{1}{\sqrt 2}\left[{\bf E}^{\rm +}_x (r, \varphi, z) + i{\bf E}^{\rm +}_y (r, \varphi, z)\right] \\
{\bf E}^{\rm -}_{\sigma^-} (r, \varphi, z)=\frac{1}{\sqrt 2}\left[{\bf E}^{\rm -}_x (r, \varphi, z) - i{\bf E}^{\rm -}_y (r, \varphi, z)\right]~.
\end{array}
\end{equation}
The red-detuned trapping field is then created by superposing these two counterpropagating modes 
\begin{equation}\label{eq:totalmode1}
{\bf E}_{\rm red} (r, \varphi, z) = \frac{1}{\sqrt 2}\left[{\bf E}^{\rm +}_{{\rm red}, \sigma^+} (r, \varphi, z)+{\bf E}^{\rm -}_{{\rm red}, \sigma^-} (r, \varphi, z)\right]~.
\end{equation}
Inserting the expressions for ${\bf E}^{\rm +}_{{\rm red}, \sigma^+}$ and ${\bf E}^{\rm -}_{{\rm red}, \sigma^-}$ from Eq.~(\ref{eq:circpolHE11}) and regrouping the terms finally yields
\begin{equation}\label{eq:totalmode2}
{\bf E}_{\rm red} (r, \varphi, z) = {\bf e}_{\beta_{\rm red} z} (r, \varphi, z) - {\bf e}_z (r) \sin(\varphi - \beta_{\rm red} z)~,
\end{equation}
where ${\bf e}_{\beta_{\rm red} z}={\bf e}_x\cos(\beta_{\rm{red}} z)+{\bf e}_y\sin(\beta_{\rm{red}} z)$ is the real-valued polarization vector field corresponding to the transverse components of a HE$_{11}$ mode, quasi-linearly polarized along a direction with azimuthal angle $\beta_{\rm red} z$. This direction of polarization thus rotates as a function of the $z$-position along the fiber with a spatial periodicity of $\lambda_{\rm red} = 2\pi/\beta_{\rm red}$, where $\lambda_{\rm red}$ is the wavelength of the red-detuned trapping light in the nanofiber. Finally, the blue-detuned field excites a quasi-circularly polarized HE$_{11}$ mode in the nanofiber waist, exhibiting an azimuthally symmetric intensity distribution which is uniform along $z$.

The combination of the red-detuned light field and the van der Waals interaction \cite{Boustimi2002} leads to an attraction of the atoms towards the fiber, whereas the blue-detuned light field repels the atoms from fiber. As the radial decay length of the evanescent field is wavelength-dependent, it thus becomes possible to create a potential minimum in the radial direction located a few hundred nanometers above the fiber surface by an appropriate choice of the optical powers. Furthermore, the azimuthal confinement originates from the azimuthal dependence of the intensity profile of the quasi-linearly polarized HE$_{11}$ modes \cite{Vetsch2010,LeKien2004b}.

Figure~\ref{Fig1}(b) shows a two-dimensional plot of the resulting light-induced potential in the $x$-$y$-plane at $z=0$ for cesium atoms. The parameters assumed in the calculation are a nanofiber diameter of 500 nm as well as a wavelength of 1064~nm (780~nm) and an optical power of $2\times 9.5$~mW (38~mW) for the red-detuned (blue-detuned) trapping field. We note that TOFs with a 500-nm diameter waist have proven to withstand such optical powers in our nanofiber-based atom trapping experiment \cite{Vetsch2010}. Figure~\ref{Fig1}(c) shows a three-dimensional view of the trapping potential. The displayed equipotential surfaces correspond to a potential energy of $k_{\rm B}\cdot 50~\mu$K above the minimum of the trap, where $k_{\rm B}$ is Boltzmann's constant and the parameters are the same as in Fig.~\ref{Fig1}(b). The light-induced potential thus corresponds to a double-helix of tubular traps. Each helix has a pitch of $h=2\pi/\beta_{\rm red}\approx 1~\mu$m and a radius of $R\approx 560$~nm. Hence, the resulting tilt angle is given by $\alpha=\arctan(h/2\pi R)\approx 16^\circ$ and the atoms are trapped at a distance of approximately 300~nm to the fiber surface. Here, we carried out calculations assuming trapping wavelengths and powers compatible with trapping cesium atoms in a deep optical potential, i.e., capable of storing atoms with a kinetic energy corresponding to the Doppler temperature. It should be pointed out, however, that our method can also be used to trap other atomic species by appropriately adapting the trapping wavelengths and the nanofiber diameter. Moreover, shallower double-helix potentials can be realized by accordingly adjusting the trapping laser powers, possibly enabling tunneling of atoms between the helices.

\begin{figure}
\centerline{\includegraphics[width=6.5cm]{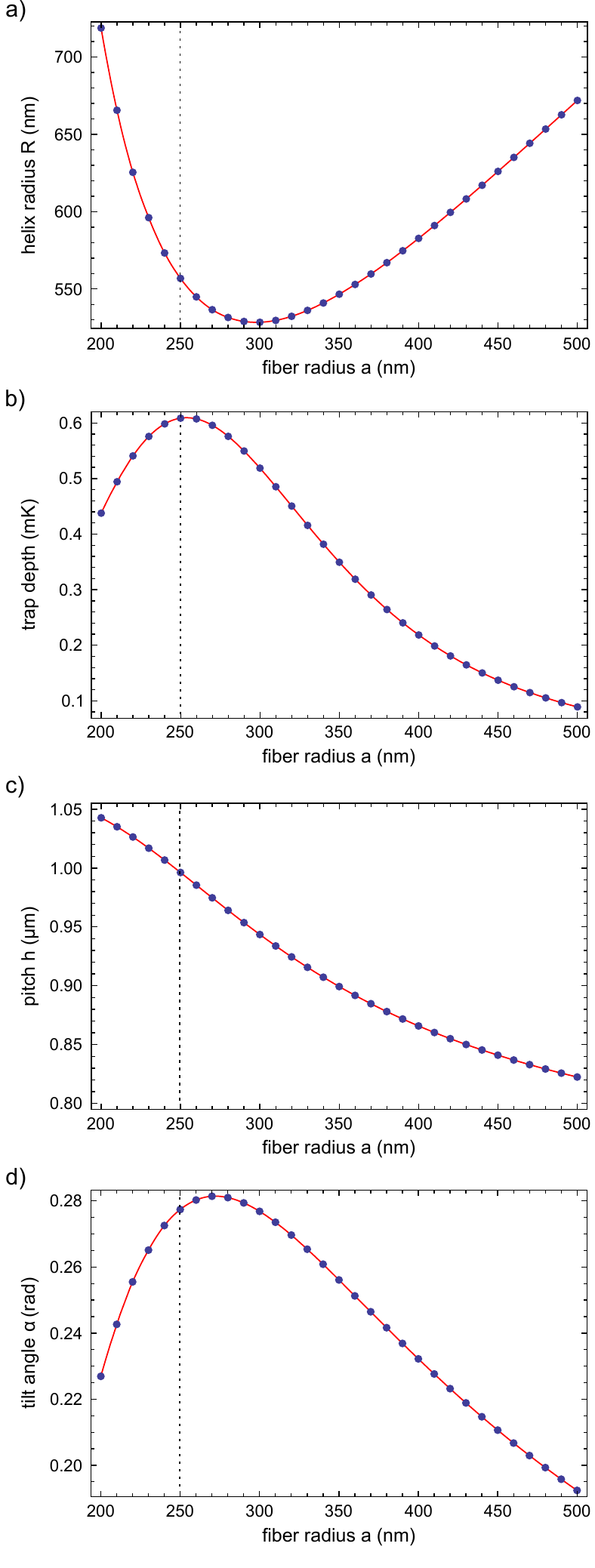}}
\caption{Dependence of the parameters of the double-helix trapping potential for cesium atoms on the radius of the nanofiber. The wavelengths and optical powers of the trapping fields are the same as in Fig.~\ref{Fig1}(b) and (c). The calculations take the van der Waals interaction into account \cite{Boustimi2002}. The solid red lines are guides to the eye.
}\label{Fig2}
\end{figure}

\begin{figure}
\centerline{\includegraphics[width=8cm]{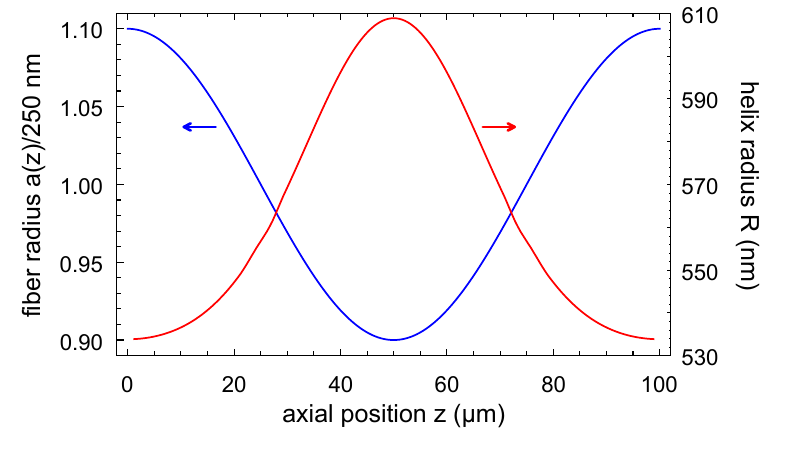}}
\caption{Example of a locally varying radius of the double helix trapping potential (red line) for a sinusoidal modulation of the nanofiber radius (blue line). The wavelengths and optical powers of the trapping fields are the same as in Fig.~\ref{Fig1}(b) and (c).
}
\label{Fig3}
\end{figure}

The nanofiber thus offers a straightforward method for creating a double-helix trapping potential for cold neutral atoms. In addition, it allows one to tailor the trapping potential by locally varying the diameter of the nanofiber. Figure~\ref{Fig2} shows the dependence of the different helix parameters on the nanofiber radius $a$ for the same wavelengths and powers of the trapping fields as in Fig.~\ref{Fig1}(b) and (c). 
The latter have been chosen such that a variation of the fiber radius around 250~nm (dashed line) results in a first order change of the radius of the double-helix, see Fig.~\ref{Fig2}(a), while the trap depth reaches its maximal value of 610~$\mu$K and therefore only changes to second order, see Fig.~\ref{Fig2}(b). In this radius range, the pitch and the tilt angle of each helix depend only weakly on the nanofiber radius, see Fig.~\ref{Fig2}(c) and (d), respectively. By modulating the latter, it is thus possible to locally alter the radius of the double-helix trapping potential while minimizing the variations of its depth. As an example, Fig.~\ref{Fig3} assumes a sinusoidal modulation of the nanofiber radius with a peak-to-peak amplitude of $\pm 25$~nm around the central value of 250~nm over a length of $50~\mu$m along the nanofiber. 
The resulting taper angle is sufficiently shallow to guarantee an adiabatic evolution of the fiber-guided light \cite{Love1986}. For this reason, we can locally describe the field by the above-mentioned HE$_{11}$ modes. The resulting radius of the double-helix ranges from $R_{\rm min}=535$~nm to $R_{\rm max}=610$~nm, showing that a significant radius variation can be obtained.

\begin{figure}
\centerline{\includegraphics[width=8cm]{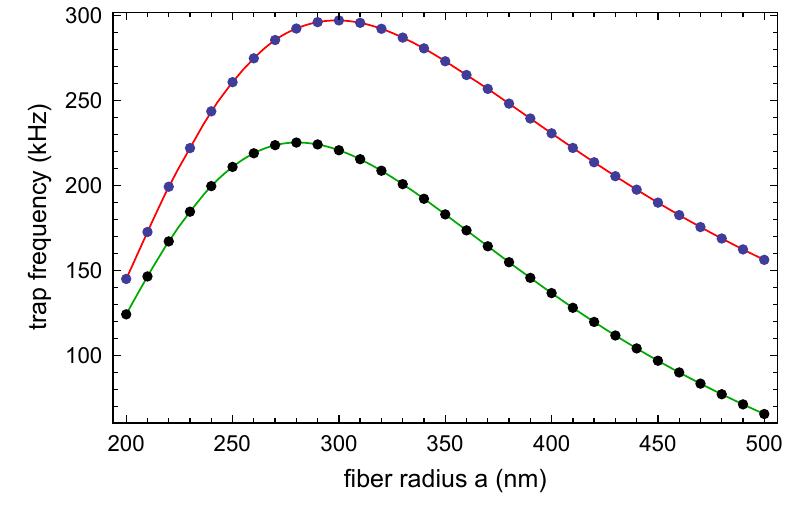}}
\caption{Dependence of the trap frequencies for cesium in the double-helix trapping potential along $v$ (black points) and $w$ (blue points) as indicated in Fig~\ref{Fig1}(c). The wavelengths and optical powers of the trapping fields are the same as in Fig.~\ref{Fig1}(b) and (c). The solid lines are guides to the eye.
}
\label{Fig4}
\end{figure}

Finally, we compute the trap frequencies for cesium transverse to the local direction $u$ of the helix along $v$ and $w$ as indicated in Fig.~\ref{Fig1}(c), see Fig.~\ref{Fig4}. Here, we assume the same parameters as in Fig.~\ref{Fig2}. For a nanofiber radius of 250~nm, both trap frequencies exceed 200~kHz. These values are high enough to remain in the one-dimensional regime even for thermal gases with temperatures of up to several $\mu$K. For a variation of the nanofiber radius of $\pm 25$~nm around 250~nm as above, the trap frequencies vary by about $\pm 30$~kHz.

Summarizing, our work shows that the use of the evanescent field of nanofiber-guided modes for the trapping of cold neutral atoms offers a flexibility in tailoring the trapping potential that is not easily accessible when using freely propagating light fields for the same purpose. In particular, it becomes possible to create a perfectly translation invariant double-helix trapping potential over the entire length of the nanofiber (typically up to $\sim 10$~mm) which is difficult to achieve with free space optics due to the divergence of propagating beams. Moreover, we have shown that a modulation of the radius profile of the nanofiber allows one to locally change the radius and pitch of the double-helix potential which would also be a challenging task in the context of free beam optics. In combination with the possibility to probe and to efficiently interface the trapped atoms using near-resonant nanofiber-guided light, nanofiber-based atom traps are thus a powerful experimental platform for fundamental research as well as for applications in quantum science and technology.

We acknowledge financial support by the Volkswagen Foundation (Lichtenberg Professorship) and the ESF (EURYI Award).

\end{document}